# The Role Specialization Model (RSM): Coordinating LLM-Based Tools in Agentic Software Development - An Exploratory Case Study

Carlos Alberto Fernández-y-Fernández [1]*, Jorge R. Aguilar Cisneros [2]

[1] Instituto de Computación, Universidad Tecnológica de la Mixteca, México

[2] Decanato de Ingenierías, UPAEP, Puebla, México

**e-mail: caff@mixteco.utm.mx*

**Abstract—**The integration of large language models (LLMs) into software development workflows has given rise to a paradigm known as Agentic Software Engineering (SE 3.0), in which autonomous agents manage full development life cycles under human supervision. This paper presents an exploratory case study in which three LLM-based tools, Antigravity (an agentic IDE with a Gemini 2.5 backend), Gemini CLI, and Qwen Code (local execution via Ollama), are coordinated according to a role-distribution framework proposed in this work as the Role Specialization Model (RSM). Three research questions guide the study: (RQ1) how can LLM-based tools with distinct capabilities be coordinated through the RSM in a real development workflow; (RQ2) what deviations from the planned role distribution emerge during RSM execution and what factors explain them; and (RQ3) how does the resulting product compare against the ISO/IEC 25010 quality model. The objective was to propose the RSM with the incremental development of a Python desktop application for interactive climate-data visualization. The workflow, observed deviations, prompt-hardening techniques, and a qualitative quality assessment are documented. Results suggest that explicit role coordination can support development cycle organization and architectural quality, but requires deliberate coordination strategies, context management, and human verification of agent-generated outputs.

## 1. Introduction

The field of software engineering is undergoing a paradigmatic transformation driven by the integration of large language models (LLMs) as autonomous agents capable of managing complete development life cycles. This evolution has been conceptualized as Agentic Software Engineering, or SE 3.0, a framework in which the human developer acts as architect, mentor, and evaluator of intelligent agents rather than as the direct producer of code [1].

In this context, the concept of vibe coding has emerged, a term popularized by Andrej Karpathy in February 2025 [2] to describe a development practice in which the programmer communicates with the model conversationally, prioritizing the perceived output over structural understanding of the generated code [3]. While this practice democratizes access to software development, it raises questions about the architectural reliability of the produced code and the long-term erosion of developers' deep technical competencies [4].

A less-explored dimension of the agentic paradigm is the possibility of coordinating multiple LLM tools with differentiated roles within a single development workflow, leveraging the specific strengths of each. This paper reports an exploratory case study in which three tools were used: (1) Antigravity, an agentic-first IDE with a Gemini 2.5 backend; (2) Gemini CLI, a command-line interface for Gemini models; and (3) Qwen Code, a code-specialized model executed locally. The study proposes and applies the Role Specialization Model (RSM), a coordination framework that assigns a distinct responsibility domain to each tool based on its capabilities.

The study is guided by three research questions: (RQ1) how can LLM-based tools with distinct capabilities be coordinated through the RSM in a real development workflow; (RQ2) what deviations from the planned role distribution emerge during RSM execution and what factors explain them; and (RQ3) how does the resulting software product compare against the ISO/IEC 25010 quality model. The contributions of this work are: (a) the RSM framework for role-specialized multi-tool LLM coordination; (b) documentation of a complete RSM-guided development workflow and its deviations; (c) identification of prompt-hardening techniques observed to support agent output integrity; (d) a qualitative product evaluation under ISO/IEC 25010; and (e) threats-to-validity analysis supporting replication efforts.

The paper is organized as follows: Section 2 reviews related work; Section 3 describes the case study design; Section 4 reports the case study execution; Section 5 presents results, discussion, sociotechnical implications, threats to validity, and a theoretical validation. Finally, Section 6 presents conclusions and future work.

## 2. Related Work

### *2.1 From Assistance to Agentic Development: The SE 3.0 Framework*

The trajectory of software engineering can be segmented into stages of incremental abstraction. Hassan et al. propose a five-level autonomy framework that characterizes this evolution (Table 1) [1]. The leap to Level 3 requires, according to the authors, a restructuring of the fundamental pillars of the discipline: actors, processes, tools, and artifacts [1]. In this new scenario, the human engineer becomes an Agent Coach, responsible for precise objective specification, execution-plan review, and artifact auditing.

Li et al. [5] analyze the emergence of AI teammates in the context of SE 3.0. They present AIDev, a large-scale dataset capturing how autonomous coding agents collaborate with humans in real-world software development. Their analysis shows that modern AI agents are capable of reading codebases, planning modifications, generating code, and submitting pull requests, thereby shifting the developer's role from manual coding to higher-level orchestration and review.

### *2.2 The Vibe Coding Phenomenon*

Vibe coding was introduced by Karpathy in February 2025 in reference to a development modality in which the programmer communicates with the model through natural language, almost without writing code directly, accepting agent outputs in a relatively passive manner [3]. The relevant academic distinction is: if the developer reviews and understands every generated line, they use an assistant; if they accept changes based on superficial results, they practice vibe coding in its strict sense [3]. Sarkar and Drosos [6] present an empirical study of vibe coding, showing that it follows an iterative goal-satisfaction model, in which developers alternate between prompting AI, evaluating generated code, and manual editing. Their outcomes suggest vibe coding does not eliminate the need for programming expertise.

This practice raises concerns about cognitive debt accumulation and the erosion of fundamental technical skills [4]. The present study addresses this risk by adopting a hybrid model in which the human developer acts as an active orchestrator, reviewing and approving outputs before integrating them into the repository.

### *2.3 Code Generation and Multi-Agent Systems*

LLM-based code generation has advanced substantially since models such as Codex [6] and AlphaCode [7] were introduced. Standardized benchmarks such as HumanEval [6] and MBPP [8] have shown that current models solve a significant proportion of introductory programming problems; however, their performance declines markedly for tasks requiring reasoning over multiple files or understanding of system architectures [9]. In the realm of multi-agent collaboration, systems such as MetaGPT [10] and ChatDev [11] assign software-team roles (analyst, designer, programmer, reviewer) to LLM instances that communicate with one another, demonstrating that role specialization improves final product quality. Unlike those fully automated systems, the present work explores a hybrid model with existing commercial and open-source tools coordinated by a human developer.

### *2.4 Instruction Following and Context Management in LLM Agents*

The effectiveness of software agents is intrinsically linked to their ability to process complex instructions while adhering to critical constraints. Recent research through the AGENTIF benchmark has identified a "constraint-density degradation" phenomenon, in which agent performance decreases as the number of simultaneous instructions increases [12]. A particularly relevant finding for this work is reported by Li et al. [13]: explicit chain-of-thought (CoT) (a prompting technique that instructs the model to reason step by step before producing output) reasoning can, in certain cases, degrade the precision of simple instruction following by redirecting the model's attention toward complex inferential processes. This effect is discussed in relation to an observed deviation in Section 5.2.

Regarding agent context management, a recent study at ETH Zurich [14] demonstrates that overly detailed project-rule files reduce task success rates and increase inference costs. The derived recommendation is to adopt failure-backed instructions: add rules only when it has been demonstrated that the agent makes a recurrent error without them.

### *2.5 Evaluation of Development Assistance Tools*

Studies on the impact of tools such as GitHub Copilot show improvements in code-writing speed, but also warn of overconfidence risks and decreased comprehension of generated code [15]. The present work contributes to this body of knowledge by documenting the interaction among multiple agentic tools throughout the complete development life cycle of a project, an aspect seldom addressed in the literature.

## 3. Case Study Design

### *3.1. Study type and research method*

This study is designed as an exploratory case study following the guidelines of Runeson and Höst [16] and Yin [17]. The case is the application of the RSM framework to a real software development project; the unit of analysis is the RSM itself, examined in terms of role adherence, deviation patterns, and quality of the resulting artifact. Data collection consisted of direct observation and systematic documentation of agent interactions, prompts, outputs, deviations, and test results during the development session. Analysis was conducted qualitatively through pattern-matching against the three research questions defined in Section 1. This study does not present a controlled experiment with independent and dependent variables; it is an exploratory observation intended to characterize and document RSM behavior in a naturalistic development setting, consistent with the "small-scale evaluation" category identified by Wohlin and Rainer [18] when a phenomenon is studied in a single, non-replicated context.

### *3.2 Tools Employed*

The experiment used three agentic tools selected to represent distinct points in the design space between reasoning capability, privacy, and usage modality.

Antigravity is an open-source IDE derived from the VS Code codebase that adopts an agentic-first approach [19]. Its distinctive feature is the native integration of an agent manager capable of autonomously executing multi-step tasks: creating files, navigating the file system, running tests, and updating code. In the configuration used, the backend was Google's Gemini 3 Flash model. It is important to note that, while some reference materials mention a "Gemini 3 Pro" model, that version was not used in our experiment; the active model was Gemini 3 in its Flash variant.

Gemini CLI is a command-line tool that enables interaction with the Gemini 3 model from the system terminal in auto mode (Gemini CLI decides the best model for the task). Its strength lies in processing tasks involving multiple files, generating extended text such as documentation, and automating pipelines through file-system output redirection.

Qwen Code is a programming-specialized language model developed by Alibaba Cloud, specialized in programming tasks [20]. In this experiment, it was executed locally via Ollama[1], ensuring data privacy by preventing transmission to external servers.

### *3.3 The Role Specialization Model (RSM)*

A central methodological contribution of this work is the Role Specialization Model (RSM), inspired by the separation-of-concerns principle [21] applied not to code but to assistive tools. The RSM proposes that each tool assumes a differentiated domain of responsibility based on its particular capabilities, so that contributions are complementary and functional overlap is minimized. The three roles assigned are described in Table 2 and Fig. 1.

This distribution is not static; as documented in Section 4.3, actual execution presented deviations that are analyzed in Section 5.2.

### *3.4 Case Study Project*

A project of moderate complexity was designed: a Python desktop application named Climate Data Visualizer, intended to load CSV files containing temperature and humidity records and represent them as interactive line charts. The project was deliberately designed to involve multiple technically relevant layers, graphical user interface design (Tkinter[2]), tabular data processing (Pandas[3]), visualization (Matplotlib[4]), input data validation, and automated unit testing, so that each RSM role would have a non-trivial task domain throughout the development cycle. See Fig. 2.

[1] https://ollama.com
[2] Tkinter: https://docs.python.org/3/library/tkinter.html

### *3.5. Quality evaluation methodology*

To address RQ3, the resulting software product was evaluated against the ISO/IEC 25010:2011 [22] product quality model. Given the exploratory, single-case nature of the study, the assessment was conducted as a qualitative expert evaluation by the first author, a method consistent with prior single-case studies of software quality. Each characteristic was operationalized as follows: functional suitability through a feature checklist review against the original requirements; performance efficiency through direct execution timing observation before and after refactoring; reliability through execution monitoring and error-log review; maintainability through static code inspection of class structure, coupling, and annotation coverage; interaction capability through an interface walkthrough; and flexibility through a cross-configuration execution test with and without the virtual environment. Ratings (High / Moderate) reflect qualitative judgments against these operationalizations, not quantitative measurements. This limitation is acknowledged in the threats-to-validity discussion (Section 5.6).

## 4. Case Study Execution

### *4.1 Phase 1: Initial Software Design with Antigravity*

The development process began with a natural-language instruction to the Antigravity agent, following a zero-shot prompting approach [23], that is, a direct task description without few-shot examples or explicit chain-of-thought directives:

[3] Pandas: https://pandas.pydata.org
[4] Matplotlib: https://matplotlib.org

*"Create a Python project to visualize climate data from a CSV file. It should have a basic graphical interface using Tkinter and allow loading the CSV and displaying line charts for temperature and humidity over time. Assume the CSV will have columns: fecha, temperatura, humedad."*

In response, Antigravity autonomously generated the directory structure, the main file climate_visualizer.py implementing the Tkinter and Matplotlib interface, support scripts (run.sh, requirements.txt), initial documentation (README.md, QUICKSTART.md), and a sample data file with 31 simulated records. A relevant finding at this stage was the agent's tendency to generate functional but monolithic code: the ClimateVisualizerApp class simultaneously handled interface logic, data loading, validation, and chart rendering, violating the separation-of-concerns principle [21]. This monolithic design identified the need for an architectural audit in subsequent phases. The result of this phase can be seen in Fig. 3.

### *4.2 Phase 2: Data Generation and Analysis with Gemini CLI*

***4.2.1. Prompt hardening.*** Generation of a 100-record test dataset via Gemini CLI revealed a critical vulnerability in instruction following: the agent attempted in its first two iterations to invoke internal tools unavailable in the restricted terminal environment. This behavior is consistent with the constraint-density degradation phenomenon documented in the AGENTIF benchmark [12]. The prompt that produced the correct result incorporated explicit negative constraints (Table 3), a prompt-hardening technique essential for preventing undesired behaviors in agents with file-system tool access [24]:

```
gemini --prompt "Generate a CSV file with 100 lines of simulated
climate data. DO NOT run commands. Do NOT use tools. DO NOT attempt
to create the file yourself. Simply write the data as plain text
here in the terminal." > weather_data.csv
```

***4.2.2. Architectural audit.*** Through Gemini CLI in interactive mode, an architectural analysis of the initial design was conducted. The analysis identified four improvement areas: (a) violation of the separation-of-concerns principle [21]; (b) inefficiency caused by recreating the Matplotlib Figure object on each update; (c) absence of type hints; and (d) overly broad exception handling. Gemini CLI implemented the proposed improvements, introducing a DataModel class to isolate data logic from the presentation layer, consistent with a Model-View-Controller decomposition.

### *4.3 Phase 3: Validation and Testing (Deviation from Original Plan)*

This phase corresponds to the most significant deviation from the original RSM plan. Under the RSM, Qwen Code was assigned the refactoring task; however, Gemini CLI assumed that task spontaneously while executing the architectural analysis of Phase 2. Qwen Code was used with a more limited scope: generating the validators.py module and the unit-test suite.

The validators.py module correctly verified the YYYY-MM-DD date format, distinguishing valid formats (e.g., 2024-01-01) from invalid ones (e.g., 2024/01/02 or 01-03-2024). The test suite generated by Qwen Code included 10 test cases across three classes (TestValidators, TestCSVValidation, TestDataModelBasic), all successful:

It should be noted that the suite covers only unit-level behavior; integration and system testing would be required before deployment claims could be made, consistent with standard software testing practice.

### *4.4 Phase 4: Advanced Interactivity with Antigravity*

The Antigravity agent implemented enhancements beyond the original requirements: a real-time statistics sidebar, interactive tooltips on data-point hover, configurable moving-average smoothing, and PNG export. Additionally, it added command-line CSV loading support. During this phase, the virtual environment became corrupted; the agent adapted by modifying run.sh to detect virtual-environment availability and fall back to system-level dependency installation.

### *4.5 Phase 5: Final Documentation with Gemini CLI*

Gemini CLI consolidated README.md and QUICKSTART.md into an improved document with a quick-start section and a five-milestone version history recording the project's evolution from the base implementation to advanced interactive features.

## 5. Results and Discussion

### *5.1 Quality Evaluation under ISO/IEC 25010 (RQ3)*

Applying the operationalizations defined in Section 3.5, Table 4 presents the qualitative assessment of the resulting product (Fig. 4) against ISO/IEC 25010:2011 [22]. Characteristic names follow the 2011 edition, which was used during the study; a future replication using the 2023 edition [25] would also address the newly added Safety characteristic, directly relevant to the security considerations in Section 5.5.

The introduction of type hints and the creation of the DataModel class were agent-driven improvements that substantially increased code analysability and testability. However, the reliability sub-characteristic was compromised by runtime environment fragility observed in Phase 4, underscoring that the quality of agentic software depends not only on the generated code but also on the robustness of the execution environment.

### *5.2 Analysis of Deviations from the Original Plan (RQ2)*

The main deviation observed was Gemini CLI assuming the architectural refactoring task originally assigned to Qwen Code. Three concurrent factors account for this: first, the absence of an explicit scope boundary between Phase 2 and Phase 3 allowed Gemini CLI to implement proposed improvements without an explicit instruction to do so. Second, functional overlap between both tools, Gemini CLI and Qwen Code, in refactoring tasks blurs the boundary between their domains. Third, workflow inertia: once Gemini CLI had proposed the improvements, the cognitive cost of switching tools outweighed the benefit of maintaining the planned distribution. Specifically, switching to Qwen Code at that point would have required re-establishing the full project context in a different tool: transferring relevant file contents, describing the work already completed, and verifying that its output would be consistent with what Gemini CLI had already analyzed. This context-switching overhead, compounded by the risk of producing inconsistent outputs across tools, constituted the cognitive cost that made continuing with Gemini CLI the pragmatic choice.

This observation is consistent with multi-agent systems literature that notes the difficulty of maintaining strict responsibility partitions when agents possess overlapping capabilities [10]. It suggests that, in real workflows, the role distribution among LLM agents tends to adapt dynamically, placing on the human orchestrator the responsibility of actively intervening when the planned distribution is deemed important.

Additionally, the Antigravity agent initially ignored the YYYY-MM-DD HH:MM date format specified in the prompt, requiring a subsequent validation phase. This behavior can be explained post-hoc by the CoT reasoning paradox documented by Li et al. [14]: when the agent focused on the broader structural reasoning task (software design generation), it deprioritized a specific formatting constraint. Importantly, this phenomenon was identified during the analysis phase of the study, not before it; the study was designed as an exploratory observation and not as a controlled experiment with pre-defined behavioral hypotheses. The fact that the observed deviation aligns with existing literature [14] is itself a finding—it provides real-world evidence of the CoT paradox in a practical agentic development context. Future work should incorporate pre-emptive mitigation strategies, such as separating format-constraint prompts from task-generation prompts, to test whether the effect can be reduced.

### *5.3 Comparative Effectiveness of the Tools (RQ1)*

Based on the documented experience, the observed strengths of each tool can be characterized within the scope of this study. Antigravity generated six project files (climate_visualizer.py, run.sh, requirements.txt, README.md, QUICKSTART.md, and a sample CSV) in Phase 1, demonstrating effective multi-file generation from a single natural-language prompt. Contextual coherence across those files (consistent column names, matching import statements, and a unified data flow) was verified by manual inspection. Gemini CLI generated the weather_data.csv dataset, produced the architectural analysis that led to the DataModel refactoring, and consolidated the project documentation in Phase 5, supporting its Analyst role. The automated pipeline in Phase 5 consisted of a single shell redirection (gemini prompt > README.md), which, while limited in scope, illustrates the tool's suitability for output-to-file automation tasks. Qwen Code generated the date-format validator and the ten unit tests within its defined Specialist scope.

### *5.4 Identified Challenges*

The experiment revealed three recurring challenges in agentic programming workflows. The first is prompt sensitivity: CSV generation required three refinement iterations with explicit negative constraints, consistent with the prompt-hardening literature [24]. The second is the need for human verification: although the code passed all unit tests, the developer intervened multiple times to correct undesired behaviors. The third is explicit context management: at several points, it was necessary to provide code snippets or project-state descriptions to guide agent actions.

A limitation of the present case study is that two of the three tools employed, Antigravity and Gemini CLI,  share the same underlying model (Gemini), which reduces the cognitive diversity of the ensemble relative to what a fully heterogeneous configuration would provide. This model homogeneity may partially explain the role overlap observed in Section 5.2, as both tools reason from the same prior. It should be noted, however, that this is a constraint of the specific instantiation reported here, not of the RSM as a framework: the model is agnostic to the backend of each tool and would equally support configurations combining models from different providers. Antigravity and Gemini CLI were selected for their accessibility and the maturity of their Gemini integration at the time of the experiment. Future work should evaluate the RSM with a heterogeneous toolset (for instance, pairing an Anthropic or Meta-based local model with a Google-backed agentic IDE) to assess whether backend diversity yields measurably different orchestration outcomes.

A particularly promising extension of the RSM is the incorporation of a fourth agent with the explicit role of evaluator, following the LLM-as-a-Judge paradigm [26]. In its original formulation, this paradigm uses a high-capability LLM as an automated judge to evaluate another model's outputs against a defined rubric, combining the scalability of automated evaluation with the nuanced reasoning of LLMs [26]. In the current RSM workflow, the audit function falls on the human developer. Partially automating this function via a judge agent could reduce orchestrator cognitive load without sacrificing product quality without sacrificing product quality, provided that known biases of the paradigm (positional, verbosity, and self-affirmation biases) are mitigated through careful rubric design and calibration against human-annotated reference outputs.

### *5.5 Sociotechnical Implications and Security*

***5.5.1 Transformation of the Developer Role.*** The SE 3.0 transition is not only a technological change but a reconfiguration of the software engineer's professional role. As Hassan et al. argue [1], the developer's value shifts from syntactic proficiency toward the ability to decompose complex problems strategically and to specify intent clearly. In the experience documented in this work, the majority of productive time was devoted not to writing code but to orchestrating the architectural transition, correcting agent deviations, and validating output integrity. This shift is reflected in Table 5.

From a critical perspective, the literature notes that agentic tools benefit experienced engineers disproportionately, enabling them to act as force multipliers [27], while developers in early career stages face difficulties in "steering" agents. This effect, which may be termed AI drag, raises concerns about the training of new engineering generations and the preservation of deep technical knowledge [4]. Future empirical studies should measure this effect longitudinally.

***5.5.2 LLMOps Production Patterns.*** The long-term stability of agent-produced software requires the adoption of LLMOps practices that treat models and prompts as first-class repository citizens [28]. The most relevant patterns identified in this study's context are: (a) prompts as code: versioning prompts in Git and applying linting to ensure output consistency; (b) separation between deterministic planning and probabilistic execution: clearly distinguishing the "what" (human-approved implementation plan) from the "how" (probabilistically generated code); and (c) intelligent model routing: delegating routine tasks to small local models (such as Qwen Code) while reserving powerful cloud models for global architectural changes. The RSM implemented in this work exemplifies the third pattern, albeit not completely deliberately, in the refactoring phase, as discussed in Section 5.2.

***5.5.3 Agentic Security: Prompt Injection and Environment Integrity.*** The autonomy granted to agents in environments like Antigravity introduces novel attack surfaces. Indirect prompt injections represent the most significant threat: malicious instructions hidden in trusted data (web pages, external CSV files) that the agent reads and executes without prior validation [24].

In the documented experiment, environment fragility manifested benignly through virtual-environment corruption and the agent's attempts to execute system commands during CSV generation. However, these incidents illustrate the escalation potential when an agent has file-system write permissions and terminal access: a successful prompt injection in a production scenario could compromise API keys or modify critical configurations [24]. This underscores the need for strict sandboxing and resource controls in agentic development environments.

### *5.6 Threats to Validity*

Following the framework of Wohlin et al. [18], the following threats are identified.

**Internal validity**. The study was conducted and analyzed by the same researcher who performed the development, introducing observer bias. Decisions about tool use, prompt design, and deviation identification were made by a single practitioner. A multi-researcher protocol or think-aloud study with an external observer would reduce this threat.

**External validity**. The study relies on a single case (one project, one developer, one toolset). Generalizability of the RSM framework and the observed deviation patterns to other projects, domains, or tool configurations cannot be assumed. In particular, two of the three tools share the same underlying model (Gemini 2.5), which may reduce cognitive diversity compared to a fully heterogeneous configuration. Replication with different tools, projects, and development teams is necessary before broader claims can be made.

**Construct validity**. The ISO/IEC 25010 evaluation was conducted as a qualitative expert assessment rather than through quantitative measurement or tool-supported analysis. The ratings reflect the first author’s judgment against the operationalizations defined in Section 3.5 and should be treated as indicative rather than definitive. Future work should operationalize the characteristics with quantitative metrics (e.g., cyclomatic complexity for maintainability, response-time measurements for performance efficiency).

**Reliability**. The study was executed once, with no replication. LLM agent behavior is probabilistic; repeating the same prompts may yield different outputs and different deviations. A replication protocol including prompt versioning and session recording would improve reliability.

### *5.7 Comparison with Existing Literature*

While empirical replication across multiple cases remains a direction for future work, the RSM can be validated theoretically by demonstrating its consistency with three independently established bodies of knowledge: the separation-of-concerns principle, multi-agent role-specialization research, and the SE 3.0 framework. This form of theoretical triangulation supports the plausibility and internal coherence of the model, following the construct validity argument recommended for exploratory frameworks in software engineering [18].

***Pillar 1: Consistency with the separation-of-concerns principle.*** The RSM is grounded in the separation-of-concerns (SoC) principle introduced by Dijkstra [21], which advocates partitioning a system so that each concern is addressed by a distinct, minimally coupled component. In the RSM, this principle is applied not to code modules but to tool roles: the Architect, Analyst, and Specialist each address a distinct concern of the development workflow (generation, analysis, and validation, respectively), with defined interfaces between them (artifact exchange) and a single integration point (the human orchestrator). The DataModel refactoring observed in Phase 2 further illustrates that applying SoC simultaneously at the code level and at the tool-orchestration level produced a more maintainable and testable artifact, consistent with the benefits predicted by the principle [21].

***Pillar 2: Alignment with multi-agent role-specialization research.*** The RSM's core coordination mechanism, assigning fixed, capability-based roles to agents, is consistent with the design rationale of two well-validated multi-agent LLM frameworks. MetaGPT [10] encodes standardized operating procedures to assign roles (Architect, Engineer, QA Engineer) to LLM instances, demonstrating that role specialization reduces hallucination cascading and improves output coherence in complex software generation tasks. ChatDev [11] organizes specialized agents into a virtual software company where each agent contributes to design, coding, or testing phases, showing that role boundaries improve end-to-end software quality. The RSM extends this principle to a hybrid human-tool setting: rather than instantiating multiple LLM agents from a single model, it coordinates distinct commercial and open-source tools selected for each role based on execution modality, capability profile, and privacy requirements. This constitutes a novel instantiation of role specialization that, as far as we know, has not been previously formalized for multi-tool agentic development workflows.

***Pillar 3: Consistency with the SE 3.0 framework.*** Hassan et al. [1] characterize Level 3 agentic software engineering as a planning–execution–verification cycle in which the human acts as Agent Coach, responsible for objective specification, plan review, and artifact auditing. The RSM operationalizes this cycle explicitly: the Architect role handles code generation (execution), the Analyst role handles analysis and documentation (planning support and architectural verification), the Specialist role handles testing and validation (verification), and the human orchestrator performs the integration and approval function described as Agent Coaching. The RSM thus functions as a concrete instantiation of the SE 3.0 model for multi-tool development workflows.

***Scope of theoretical validation.*** Theoretical validation establishes the plausibility and internal consistency of the RSM but cannot substitute for empirical evidence across multiple cases. The alignment with SoC [18], MetaGPT [10], ChatDev [11], and SE 3.0 [1] demonstrates that the RSM's principles are principled extensions of validated concepts to a new context rather than ad-hoc design decisions. Empirical validation through replication, varying projects, toolsets, and developers, remains the necessary next step, as outlined in Section 6.

## 6. Conclusions

This paper presented an exploratory case study of the Role Specialization Model (RSM), a coordination framework for multi-tool LLM-based development workflows, applied to the development of a climate-data visualization application in Python.

Addressing RQ1, the RSM provided a structured basis for tool coordination; however, actual execution deviated from the planned role distribution due to functional overlap and workflow inertia, placing responsibility for actively managing these deviations on the human orchestrator. Addressing RQ2, three concurrent factors explained the observed deviation: absent scope boundaries, tool capability overlap, and the cognitive cost of context-switching. Addressing RQ3, the qualitative ISO/IEC 25010 assessment indicated high functional suitability, maintainability, and flexibility, with moderate reliability due to environmental fragility.

The main conclusions are: (a) role specialization is a valid design principle for multi-tool LLM coordination but requires explicit scope boundaries to prevent unintended overlaps; (b) prompt hardening through explicit negative constraints was found to be a critical technique in this study for ensuring agent output integrity in environments with system-tool access, though further cases are needed to confirm generalizability; (c) human verification of agent-generated outputs remains indispensable; (d) the CoT reasoning paradox [14] and constraint-density degradation [13] were observed in naturalistic conditions, contributing real-world evidence to these phenomena.

A particularly promising extension of the RSM is the incorporation of a fourth agent with the explicit role of evaluator, following the LLM-as-a-Judge paradigm [26]. In the current RSM workflow, the audit function falls to the human developer; a judge agent could reduce orchestrator cognitive load, provided that known biases of the paradigm (positional, verbosity, and self-affirmation biases) [26] are mitigated through careful rubric design. Future work should also evaluate the RSM with a heterogeneous toolset—for instance, pairing an Anthropic or Meta-based local model with a Google-backed agentic IDE—to assess whether backend diversity yields measurably different coordination outcomes. Additional directions include: extending the study to larger-scale projects; developing quantitative orchestration effectiveness metrics, including exploration of LLM-as-a-Judge as an integrated evaluation mechanism; and longitudinally evaluating the impact of agentic tools on technical competencies of developers at different career stages.

**Declaration of Competing Interests**

The authors declare that they have no known competing financial interests or personal relationships that could have appeared to influence the work reported in this paper.

**CRediT Author Contribution Statement**

C.A. Fernández-y-Fernández: Conceptualization, Methodology, Software, Investigation, Writing - Original Draft, Writing - Review & Editing. Jorge R Aguilar-Cisneros: Writing - Review & Editing.

**Data Availability**

The code supporting this case study are available from the corresponding author upon reasonable request.

## Appendix A. Generated Code

The code generated by the agents using the RSM framework is available as a Git repository at 10.5281/zenodo.21076890 [5].

## Tables

**Table 1.** Levels of autonomy in AI-assisted software engineering. Adapted from [1]

| Level | Designation | Unit of work | Workflow characteristics |
|---|---|---|---|
| 1 | Token assistance | Tokens / lines | Inline suggestions based on cursor position (SE 1.5) |
| 2 | Task assistance | Functions / files | Code block generation in a single attempt (SE 2.0) |
| 3 | Objective-agentic | Features / tickets | Planning–execution–verification cycles (SE 3.0) |
| 4 | Domain autonomy | Modules / services | Self-management within specific technical domains |
| 5 | General autonomy | Complete systems | Minimal human intervention; the agent acts as a professional peer |

[5] When the article is published, a link to a public repository containing these and additional files will be provided to support reproducibility. Available to reviewers from the corresponding author upon request.

**Table 2.** Role distribution among tools in the Role Specialization Model (RSM)

| Tool | Assigned role | Execution mode | Primary strength |
|---|---|---|---|
| Antigravity | Architect | Agentic IDE (cloud) | Multi-file management, macro vision, iterative refinement |
| Gemini CLI | Analyst | Terminal (cloud) | Bulk processing, documentation generation, and architectural audit |
| Qwen Code | Specialist | Local (Ollama) | Data validation, unit testing, sensitive-data privacy |

**Table 3.** Prompt-hardening iterations for CSV generation via Gemini CLI

| # | Prompting strategy | Technical outcome |
|---|---|---|
| 1 | Direct CSV generation request | Error: agent attempted to invoke unavailable tools |
| 2 | Terminal output redirection instruction | Partial failure: mix of metadata and plain text in output |
| 3 | Inclusion of explicit negative constraints | Success: clean data generation redirected to file |

**Table 4.** Qualitative software quality evaluation under ISO/IEC 25010:2011. Assessment methods defined in Section 3.5. Characteristic names reflect the 2011 edition; the 2023 edition renames Usability to Interaction Capability and Portability to Flexibility.

| Characteristic | Sub-characteristic | Assessment method | Observed result |
|---|---|---|---|
| Functional suitability | Completeness | Feature checklist review | High: all planned features implemented (charts, filters, export, statistics panel) |
| Performance efficiency | Time behaviour | Execution timing observation | Improved after refactoring: Figure/Canvas reuse eliminates latency on data reload |
| Reliability | Faultlessness | Execution and error-log review | Moderate: sensitive to runtime environment stability (venv corruption observed) |
| Maintainability | Modularity / analysability | Static code inspection | High: MVC structure, type hints throughout, centralised column-name constants |
| Interaction capability | User engagement | Interface walkthrough | High: clean layout, standard Matplotlib navigation toolbar, interactive tooltips |
| Flexibility | Adaptability | Cross-configuration test | High: run.sh detects the environment automatically and applies a fallback installation |

**Table 5.** Shift in professional value metrics in the SE 1.0–2.0 → SE 3.0 transition. Adapted from [1] and illustrated by observations from this study.

| Traditional values (SE 1.0/2.0) | Agentic values (SE 3.0) |
|---|---|
| Manual algorithm writing | System architecture and design |
| Syntax debugging | Agent orchestration and output review |
| Typing speed | Specification clarity and intentionality |
| Manual unit testing | Audit of agentic test plans |

## Figure Captions

**Fig. 1.** RSM architecture depicting the agent's role and human orchestrator.

**Fig. 2.** Instantiation of the RSM for the Climate Data Visualizer case study, showing tool assignment, phase distribution, and the unplanned role deviation in Phase 3.

**Fig. 3.** Application loaded with sample_climate_data.csv (31-day dataset). Initial version without statistics panel or smoothing, produced by Antigravity in Phase 1.

**Fig. 4.** Climate Data Visualizer application after loading weather_data.csv. The left panel shows real-time statistics; the main panel displays dual temperature (red) and humidity (teal) line charts with a navigation toolbar.

## Figures

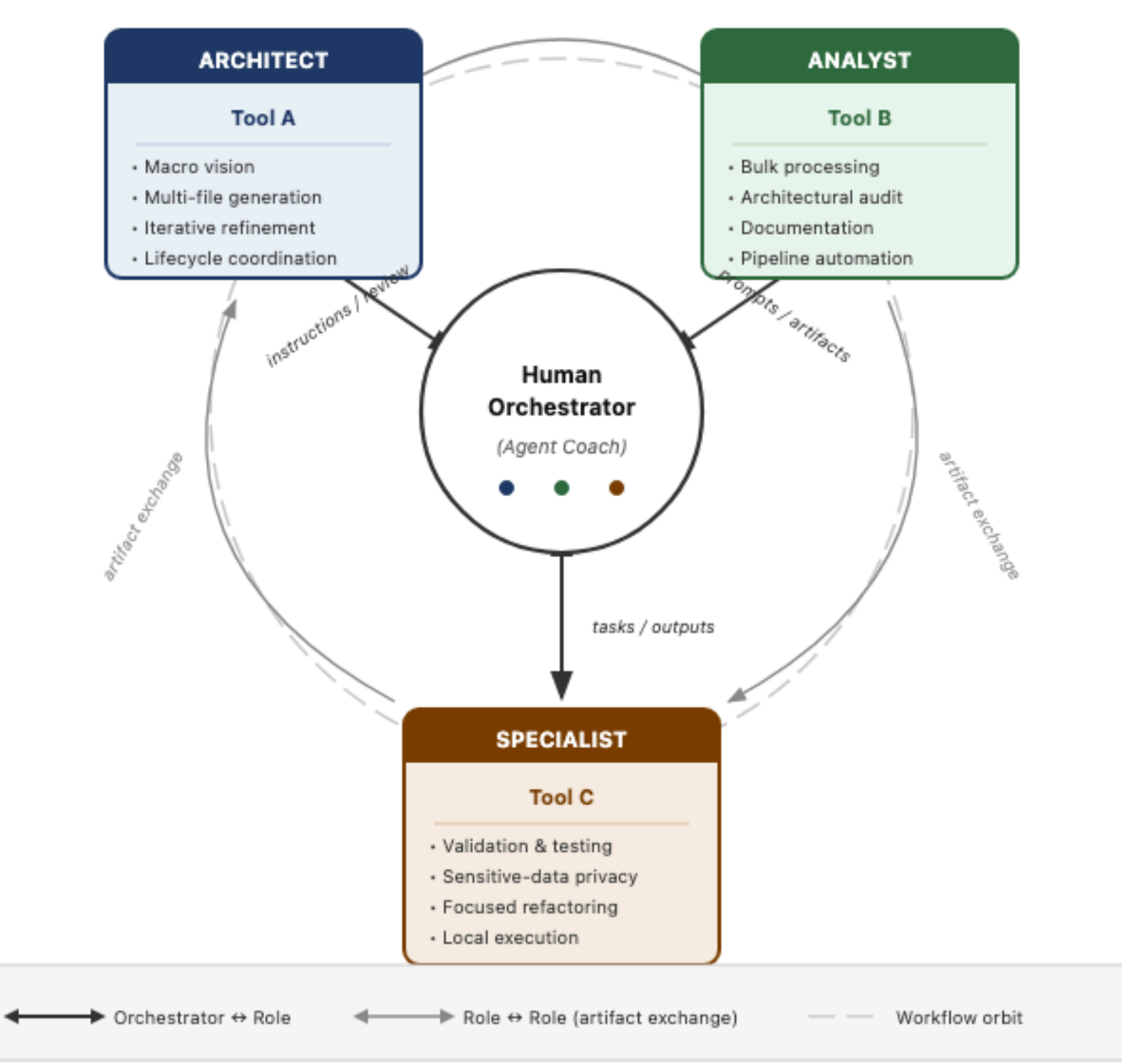

Role Specialization Model (RSM)
A multi-agent LLM orchestration framework for software development
artifact exchange
ARCHITECT
Tool A
• Macro vision
• Multi-file generation
• Iterative refinement
• Lifecycle coordination
ANALYST
Tool B
• Bulk processing
• Architectural audit
• Documentation
• Pipeline automation
instructions / review
prompts / artifacts
Human
Orchestrator
(Agent Coach)
artifact exchange
artifact exchange
tasks / outputs
SPECIALIST
Tool C
• Validation & testing
• Sensitive-data privacy
• Focused refactoring
• Local execution
Orchestrator ↔ Role
Role ↔ Role (artifact exchange)
Workflow orbit

Fig. 1.

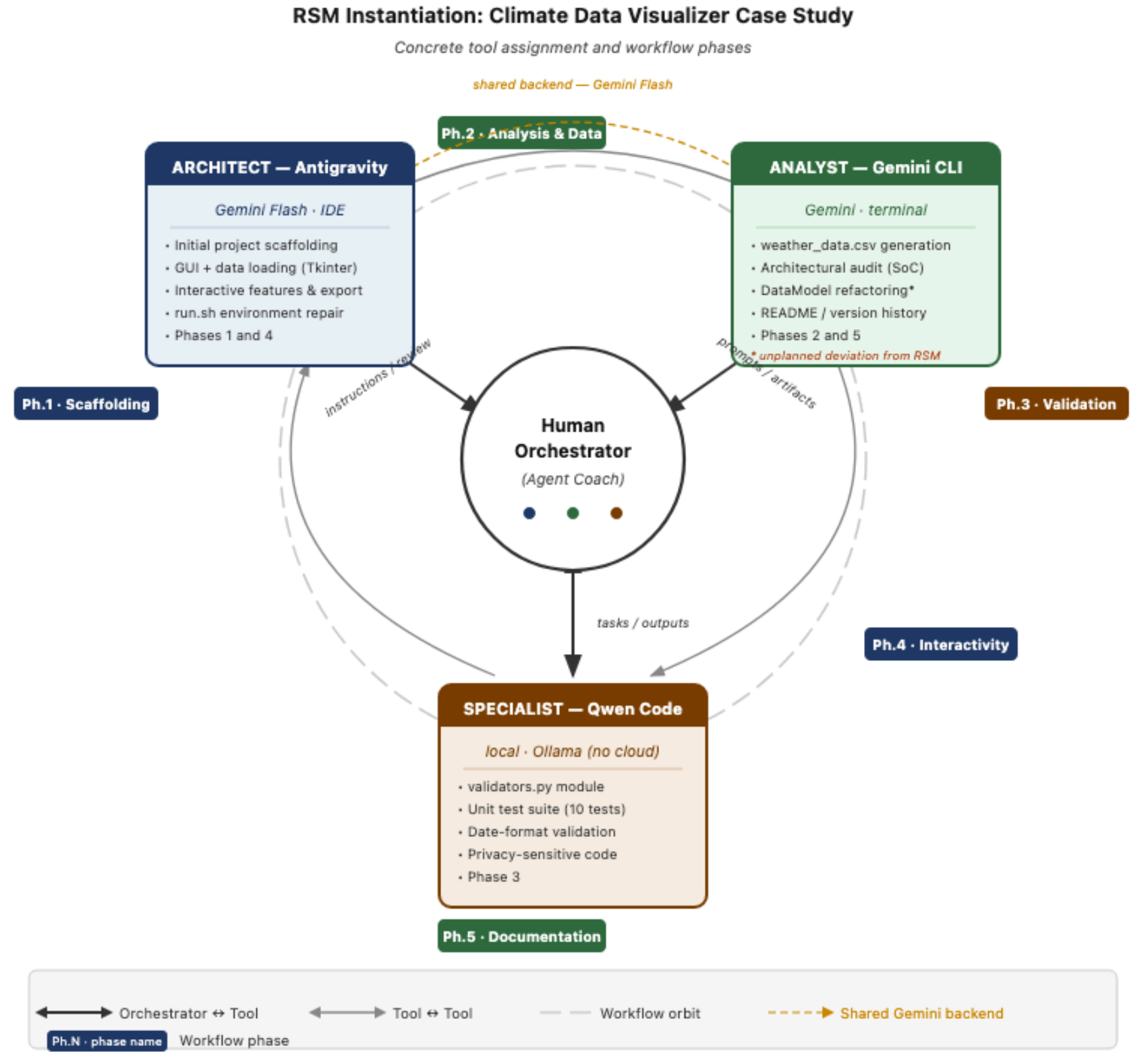

RSM Instantiation: Climate Data Visualizer Case Study
Concrete tool assignment and workflow phases
shared backend — Gemini Flash
Ph.2 · Analysis & Data
ARCHITECT — Antigravity
Gemini Flash · IDE
• Initial project scaffolding
• GUI + data loading (Tkinter)
• Interactive features & export
• run.sh environment repair
• Phases 1 and 4
ANALYST — Gemini CLI
Gemini · terminal
• weather_data.csv generation
• Architectural audit (SoC)
• DataModel refactoring*
• README / version history
• Phases 2 and 5
* unplanned deviation from RSM
Ph.1 · Scaffolding
Ph.3 · Validation
instructions / review
Human
Orchestrator
(Agent Coach)
tasks / outputs
Ph.4 · Interactivity
SPECIALIST — Qwen Code
local · Ollama (no cloud)
• validators.py module
• Unit test suite (10 tests)
• Date-format validation
• Privacy-sensitive code
• Phase 3
Ph.5 · Documentation
Orchestrator ↔ Tool
Tool ↔ Tool
Workflow orbit
Shared Gemini backend
Ph.N · phase name
Workflow phase

Fig. 2.

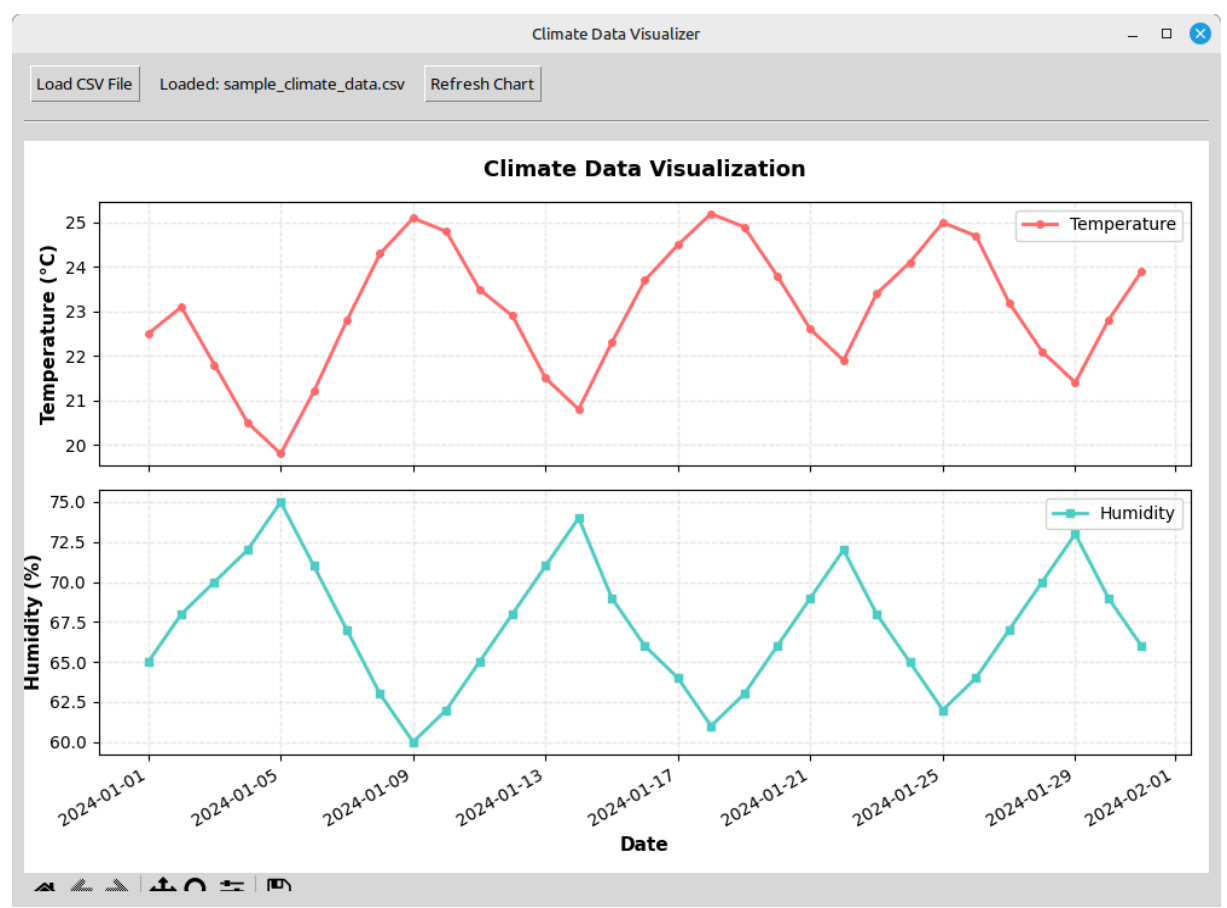

Fig. 3.

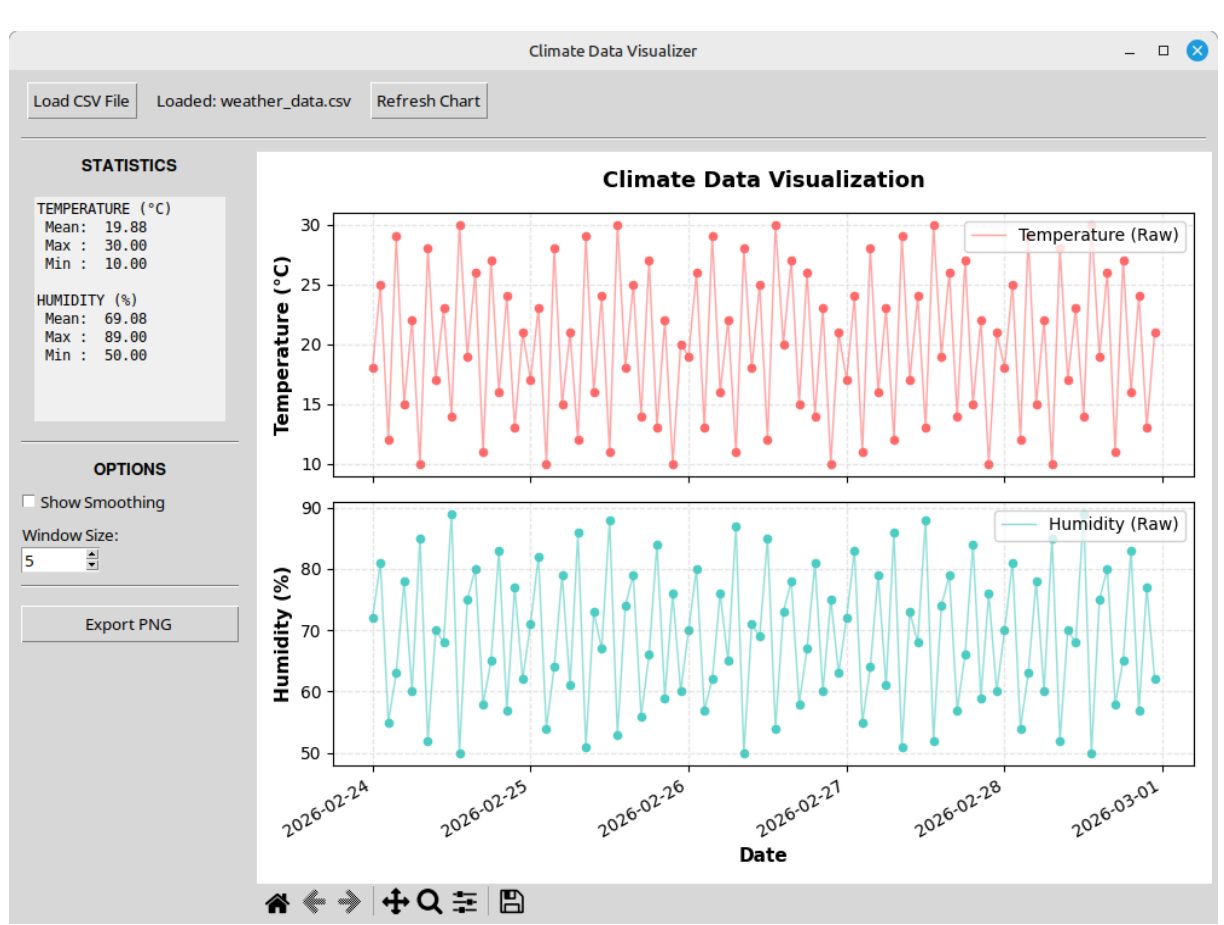

Fig. 4.